\documentclass[number,sort&compress]{elsarticle}

\usepackage{amsmath,hyperref,amssymb}
\usepackage[hyperref]{xcolor}
\hypersetup{
	colorlinks = true}
\journal{Nuclear Instruments and Methods in Physics Research Section A}

\begin{document}
\begin{frontmatter}
\title{A HEMT-Based Cryogenic Charge Amplifier with sub-100 eVee Ionization Resolution for Massive Semiconductor Dark Matter Detectors} 

\author[stanford]{A.~Phipps\corref{cor1}}
\ead{arran@stanford.edu}
\author[lyon]{A.~Juillard}
\author[berkeley]{B.~Sadoulet}
\author[berkeley]{B.~Serfass}
\author[cnrs]{Y.~Jin}
\cortext[cor1]{Corresponding author}
\address[stanford]{Department of Physics, Stanford University, Stanford, CA 94305, USA}
\address[lyon]{Universit\'{e} de Lyon, Universit\'{e} Lyon 1, CNRS/IN2P3, IPN-Lyon, F-69622 Villeurbanne Cedex, France}
\address[berkeley]{Department of Physics, University of California, Berkeley, CA 94720, USA}
\address[cnrs]{C2N, CNRS, Univ. Paris-Sud, Univ. Paris-Saclay, 91460 Marcoussis, France}

\begin{abstract}
We present the measured baseline ionization resolution of a HEMT-based cryogenic charge amplifier coupled to a CDMS-II detector. The amplifier has been developed to allow massive semiconductor dark matter detectors to retain background discrimination at the low recoil energies produced by low-mass WIMPs. We find a calibrated baseline ionization resolution of $\sigma_E = 91\,\text{eV}_{ee}$. To our knowledge, this is the best direct ionization resolution achieved with such massive ($\approx$150 pF capacitance) radiation detectors.

\end{abstract}


\end{frontmatter}
\section{Introduction}
Semiconductor radiation detectors are sensitive to recoil events caused by scattering dark matter in the form of Weakly Interacting Massive Particles (WIMPs). In the $\sim$1-100 GeV/$c^2$ mass range, most theories predict that WIMP scatters will preferentially induce nuclear recoils in the target material, often made from high-purity germanium or silicon crystals. The scattering of a WIMP would be a rare event. In these detectors, most events are caused by electromagnetic background which preferentially scatter off electrons. The Cryogenic Dark Matter Search (CDMS)  \cite{cdms2000, cdms2005, cdms2009, cdms2010} and EDELWEISS experiments  \cite{EDW_LTD16,EDW_2011,EDW_2017, EDW_2016} have a history of setting competitive WIMP detection limits by rejecting background events through the simultaneous measurement of the produced ionization and phonons. The ionization yield (the ratio of produced ionization to phonons) is smaller for nuclear recoils and allows these detector to discriminate between signal and background on an event-by-event basis.

Historically, there has been a preference for direct detection searches to focus sensitivity on WIMP masses $\sim$10 GeV/$c^2$ and higher, producing keV-scale recoils. This mass range is well-suited to the JFET-based charge amplifiers traditionally used for ionization readout, which have a typical baseline ionization resolution of $\sigma_E = 0.4 \,\text{keV}_{ee}$\footnote{$\text{keV}_{ee}$ stands for keV-electron-equivalent, the amount of ionization energy produced by a 1 keV electronic recoil event.}. This range has been extensively searched with no credible detections. The remaining parameter space will be well-probed by higher-mass liquid noble detectors such as LZ \cite{lz} and XENON \cite{xenon1t}, which benefit from improved kinematics, greater total exposure to high-mass WIMPs.

Cryogenic semiconductor detectors have instead shifted focus to low-mass WIMPs, where the lower atomic mass of germanium and silicon is favored  \cite{cogent2013lowmass,cdmsreplace,EDW_2018_optimize}. The increase in low-mass WIMP count rate with decreasing baseline energy resolution has motivated a number of novel techniques to lower thresholds, including single electron-hole pair detection \cite{sensei,damiclowmass,cdms2018ehpair}. These techniques, however, lack the ability to distinguish between electron and nuclear recoils through ionization yield. This has driven the CDMS and EDELWEISS collaborations to pursue the development of charge amplifiers for ionization readout with a target baseline ionization resolution of $\sigma_E$ = 100 eV$_{ee}$ or better. In this work, we demonstrate a calibrated baseline ionization resolution of $\sigma_E = 91\,\text{eV}_{ee}$ in a CDMS-II detector coupled to a HEMT-based cryogenic charge amplifier of our own design.  
\section{HEMT-based Charge Amplifier}
Charge amplifiers integrate the drift current generated within the cryogenic semiconductor detector by an interacting particle and output a pulse whose amplitude is proportional to the total charge produced. A high impedance input stage with a capacitance well-matched to the detector capacitance of $C_{det}\approx 150$ pF is required to maximize the signal-to-noise. Dominant sources of noise are the current and voltage noise of the input stage and the Johnson noise from amplifier feedback and detector bias resistors.

High electron mobility transistors (HEMTs) have been developed by {CNRS-C2N}  \cite{YongAPL} to replace JFETs as the primary input component for the readout of cryogenic semiconductor radiation detectors. Compared to equivalent JFETs, which work best at T=120 K, the HEMTs have better noise performance, 50$\times$ lower power dissipation (100 $\mu$W vs 5 mW per transistor) and can operate at liquid helium temperatures without freezing out. We have previously demonstrated equivalent ionization performance with a HEMT-based input coupled to room-temperature CDMS-II electronics  \cite{CDMSwHEMT}. By using multiple HEMTs, we have developed a fully-cryogenic charge amplifier with improved noise performance, described in  \cite{LTD16HEMTpaper}.

\begin{figure}
\centering
\includegraphics[width=\linewidth]{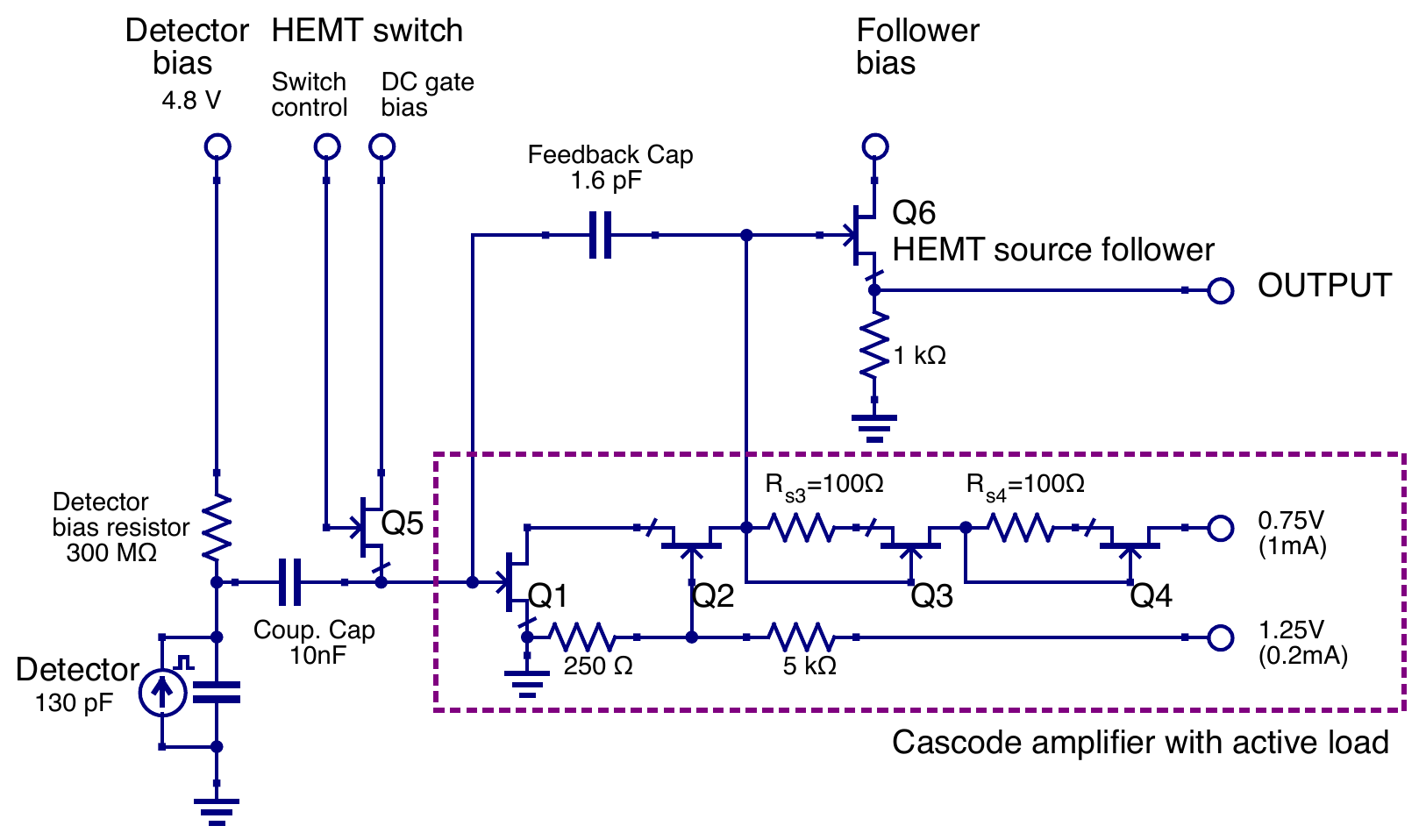}

%
\caption{Schematic of the HEMT-based charge amplifier. The dashed box surrounds the cascode and active load, which form the open-loop voltage amplifier. The cryogenic semiconductor detector is AC-coupled to the input. Closed-loop charge amplification is achieved by a small feedback capacitor placed between the open-loop input and output. The DC input voltage is set using a HEMT switch, which remains open during standard operation. A HEMT source-follower isolates the amplifier output from high capacitance cabling to room temperature.}
\label{fig:fullAmp}
\end{figure}

The design of the HEMT-based charge amplifier is shown on Fig.~\ref{fig:fullAmp}.
It uses six HEMTs (Q1-Q6) with three types of gate geometries, developed by {CNRS-C2N}  \cite{YongAPL}. The gate areas are 32 $\mu$m x 2 mm for Q1, 4 $\mu$m x 5 mm for Q2-Q4, and 2 $\mu$m x 1 mm for Q5-Q6. All components are located on a custom-designed PCB thermalized to the 4K stage of the cryostat. Power is provided by room-temperature supplies filtered by 4 mF capacitors. The total power dissipation for the entire amplifier is only 1 mW, significantly less than the 5 mW dissipated by a single JFET. 

The core amplifier is a cascode  \cite{Gray} that converts the signal voltage at the input of Q1 into a current via its transconductance with a boosted output impedance due to the addition of Q2. The cascode properties allows it to drive this current through the active load formed by Q3 and Q4, generating an inverted, amplified voltage signal at the output. This forms the open-loop voltage amplifier which had a measured voltage gain of 340.

Charge amplification is achieved by connecting a small 1.6 pF feedback capacitor $C_{fb}$ between the input and output. The cryogenic semiconductor detector, which must be biased at several volts for charge collection, is AC-coupled to the amplifier input through a 10 nF capacitor. Charge collection in the detector induces a parallel current at input of the amplifier which is integrated onto the feedback capacitor, producing an increase in the output voltage proportional to the total charge. The total charge gain depends on the feedback capacitance, detector capacitance, and open-loop voltage gain. The gain is calibrated in units of keV$_{ee}$ by exposing the detector to known radioactive sources and identifying the corresponding photopeaks.

To eliminate its Johnson noise contribution, the usual parallel feedback resistor used to passively reset the feedback capacitor is omitted. Instead, a HEMT switch (Q5) is used to actively reset the amplifier  \cite{HEMTswitch}.
The switch periodically resets the DC bias voltage at the amplifier input ($V_{\mathbf{DC}}$=-25 mV) and remains open during data acquisition. The switch HEMT has a small input capacitance of 5 pF to limit charge injection while switching and capacitive noise injection when open. The switching transients are short (10 $\mu$s) and leakage currents with the switch open are negligible. The amplifier output is buffered from loading due to room-temperature cabling through a cryogenic HEMT source-follower (Q6).

Measurements of the intrinsic noise performance of the amplifier were previously reported in  \cite{LTD16HEMTpaper}. In conjunction with single-HEMT  measurements performed at CNRS-C2N, an input-referred noise model of the amplifier has been developed  \cite{YongAPL,EDWreadout}. The input voltage noise consists of a white noise contribution and a $1/f$ contribution, expressed as
\begin{equation}
\label{eq:ennoise}
e_{n}=\sqrt{(0.23\cdot10^{-9})^2+(10\cdot 10^{-9}/\sqrt{f})^2}~~[\text{V}/\sqrt{\text{Hz}}].
\end{equation}
The current noise is described by
\begin{equation}
\label{eq:innoise}
\begin{split}
&i_{n}=\sqrt{i_{0}^2+a^2\cdot f}~~[\text{A}/\sqrt{\text{Hz}}]~~\text{with}\\
&i_{0}=\sqrt{2\cdot q\cdot I_{leak}}~~[\text{A}/\sqrt{\text{Hz}}],\,a=2\cdot 10^{-17}~~[\text{A/Hz}],
\end{split}
\end{equation}
where $I_{leak}$ has been measured to be less than $3\times 10^{-17}$ A for the amplifier itself. The total voltage noise at the input will depend on the total input impedance $Z_{inTot}$ and is given by
\begin{equation}
\label{eq:entot}
\begin{split}
e_{nTot}&=\sqrt{e_{n}^2+(Z_{in~Tot}\cdot i_{n})^2}~~[\text{V}/\sqrt{\text{Hz}}],~\text{with}\\
Z_{inTot}&=1/(2\cdot \pi \cdot f\cdot (C_{in}+C_{fb}))~~[\Omega]\\
\end{split}
\end{equation}
in the absence of a detector bias resistor, where $C_{in}$ is the sum of the HEMT gate capacitance (100 pF), detector capacitance, and cabling capacitance. The measured noise performance is very close to that of a single HEMT, confirming the total amplifier noise is completely dominated by the noise of the input HEMT Q1.

\section{Performance with a CDMS Detector}
To study the ionization resolution, the HEMT amplifier was connected to a CDMS-II  \cite{cdms2000} germanium detector. The detector was 1 cm thick, 7.6 cm in diameter with a mass of 240 g. The detector center electrode and outer guard ring, typically read out by separate charge amplifiers to reject events occurring near the detector sidewalls, were connected in parallel to accommodate the single-channel HEMT amplifier prototype. The total capacitance of the detector was estimated to be 130 pF, giving a total input capacitance (including cabling) of 240 pF. The electrodes were biased at 4.8 V through a 300 M$\Omega$ bias resistor placed in close proximity to the detector. Both the bias resistor and detector were thermally sunk to the 40 mK base temperature stage of the cryostat. Similar to previous work  \cite{CDMSwHEMT}, the detector was exposed to a collimated $^{241}$Am source facing the detector to provide a calibrated measurement of the ionization resolution. Taking into account the total input impedance and additional Johnson noise of the bias resistor, the noise model from Eq.~\ref{eq:entot} predicts a baseline ionization resolution of $103\,\text{eV}_{ee}$. 

\begin{figure}
\centering
\includegraphics[%
  width=0.49\linewidth,
  keepaspectratio]{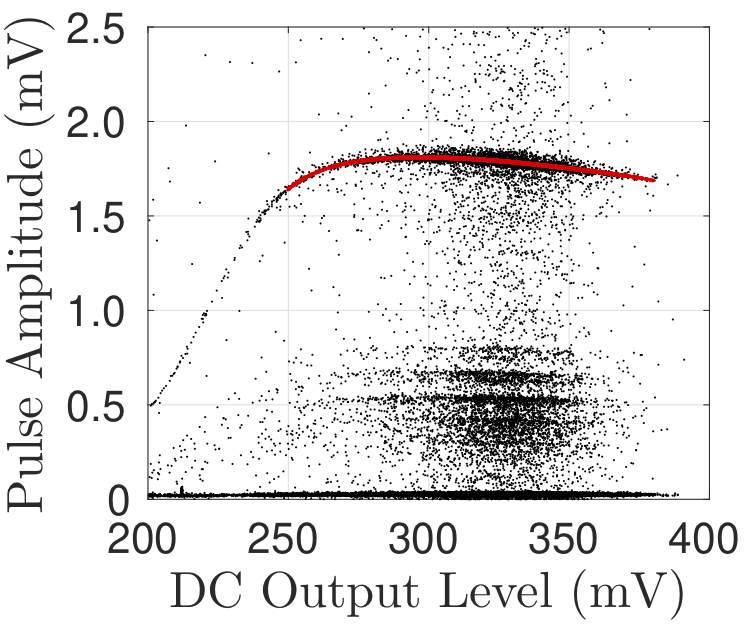}
\includegraphics[%
  width=0.49\linewidth,
  keepaspectratio]{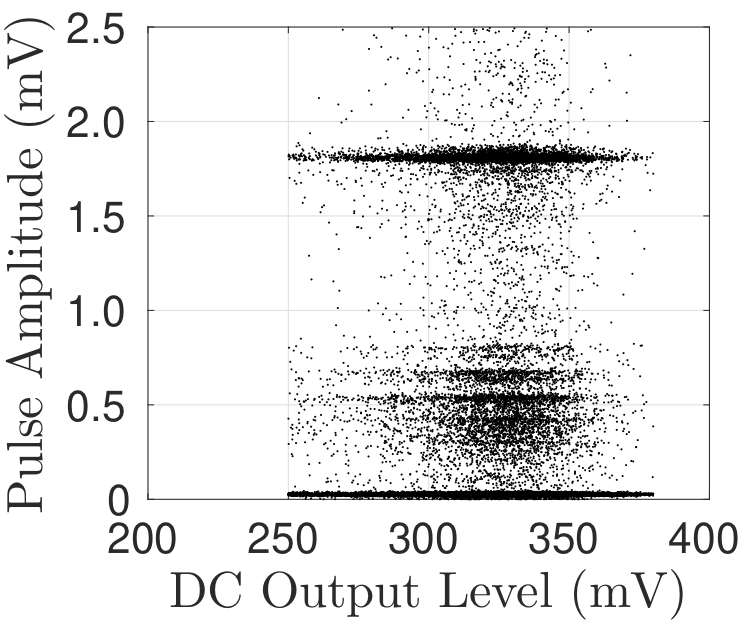}
  \caption{Pulse amplitude (in mV) of the $^{241}$Am data vs HEMT amplifier follower DC output. {\it Left:} Uncorrected data. The red line shows the polynomial fit to the 59.5 keV photopeak used to define the gain correction (see text). {\it Right:} After correction.}
\label{fig:ampVsFO}
\end{figure}

Pulse data from the HEMT amplifier follower output was sampled at 200 kHz with a 16-bit ADC. An internal 5-pole Bessel filter at 100 kHz was used to prevent aliasing. For each pulse, the follower DC level was recorded. After application of a first-order 100 Hz high pass filter, an optimal filter  \cite{OFreso} frequency domain fit determined the pulse amplitudes and start times. Both were found to be in agreement with time domain fits performed on the same data.

We observed some variation in the amplifier gain due to a combination of the high event rate at the surface test facility and low open-loop gain of the amplifier. Recoil events with energies of several MeV (from alphas, high energy muons, or pileup events) would deposit enough charge to shift the DC level of the amplifier input and modulate the gain between the periodic resets performed by the switch. The DC input level controls the voltage present on the gate of the follower (Q6), which determines the follower DC output level. The follower DC output level thus tracks the accumulated charge at the input (Q1) gate. As seen in Fig.~\ref{fig:ampVsFO}, the gain variation tracked with follower DC output level and was corrected on an event-by-event basis. An initial 250-380 mV follower DC output selection cut was applied to reject events with significant ($\geq 10\%$) gain variation. The gain correction was then defined by a 5$^{th}$ order polynomial fit (red line) to the 59.5 keV photopeak. Although this gain variation would not be expected to occur during low-background WIMP search conditions, we note it can be reduced by increasing the value of the feedback capacitor (reducing the charge amplification) or increasing the open loop voltage gain of the HEMT amplifier. 

\begin{figure}
\centering
\includegraphics[%
  width=0.49\linewidth,
  keepaspectratio]{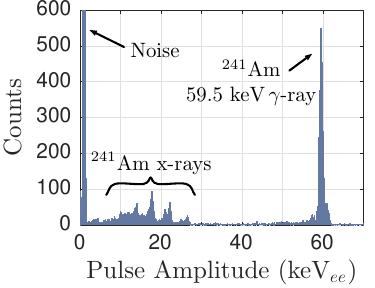}
    \includegraphics[%
  width=0.49\linewidth,
  keepaspectratio]{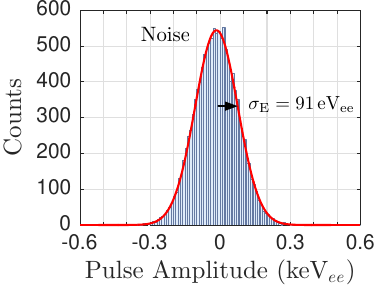}
\caption{{\it Left:} Histogram of the $^{241}$Am data in $\text{keV}_{ee}$, with a bin width of 0.1 $\text{keV}_{ee}$. The main lines of $^{241}$Am $L$ X-rays are 13.9, 17.8 and 20.8 keV and $\gamma$-rays are 26.4 and 59.5 keV. {\it Right:} Histogram of the amplitude distribution produced by applying the frequency domain optimal filter fit to a large number of noise traces. The RMS baseline ionization resolution deduced from a Gaussian fit (red curve) is $\sigma_E = 91\,\text{eV}_{ee}$.}
\label{fig:histoAm}
\end{figure}

Fig.~\ref{fig:histoAm}({\it Left}) shows the resulting histogram of the gain-corrected data of Fig.~\ref{fig:ampVsFO}, calibrated to $\text{keV}_{ee}$ based on the location of the 59.5 keV photopeak. Photopeaks from lower energy $L$ X-rays are also visible at the correct locations, indicating good linearity over this recoil energy range. The RMS resolution of the 59.5 keV peak was 420$\pm15$ $\text{eV}_{ee}$, in agreement with the resolution measured with the CDMS JFET amplifier (440 $\text{eV}_{ee}$)  \cite{CDMSwHEMT} coupled to the same detector during a separate cryogenic run. At these energies, the photopeak resolution is dominated by detector effects including the statistics of electron-hole pair creation and charge trapping  \cite{trapping1,trapping2} and is not indicative of the baseline ionization resolution. 

The baseline RMS ionization resolution was found by applying the frequency domain optimal filter pulse fitting routine to a large number of noise traces in which no recoil events are present. Following standard practice, the pulse start time was held constant to prevent the fit from always identifying the maximum noise fluctuation as the pulse amplitude, which would produce a shifted non-Gaussian amplitude distribution and an incorrect result. The baseline ionization resolution is defined by the width of the resulting amplitude distribution. As shown in Fig.~\ref{fig:histoAm} ({\it Right}), the frequency domain optimal filter obtained an RMS resolution of $\sigma_E = 91\,\text{eV}_{ee}$, compared to $\sigma_E =130\,\text{eV}_{ee}$ when using the time domain fit (not shown). The measured resolution is $\approx$10\% better than the predictions of Eq.~\ref{eq:entot}, possibly due to errors in the estimated closed-loop gain used for the calculation (27 nV/eV$_{\text{ee}}$ predicted vs 30 nV/eV$_{\text{ee}}$ measured) or detector-related effects such as impact ionization  \cite{impact}, demonstrating the importance of using radioactive sources for an unbiased calibration. To our knowledge, this is the best direct baseline ionization resolution achieved with such massive semiconductor detectors of $\approx$150 pF capacitance.            

\section{Conclusions}
In conclusion, we have shown that sub-100 $\text{eV}_{ee}$ baseline ionization resolution is achievable in massive dark matter detectors such as those used by the CDMS and EDELWEISS collaborations. Our cryogenic HEMT-based charge amplifier is a significant advancement over JFET-based designs, providing improved ionization resolution with lower power dissipation and elimination of the thermal standoffs required to prevent JFET freeze-out. While better gain stability is desirable for high-background conditions, the use of a CDMS-II detector in this work demonstrates the feasibility of this design for future CDMS and EDELWEISS low-background dark matter searches. The retention of background discrimination through ionization yield at low recoil energies, enabled by the improved baseline ionization resolution, allows for increased exposure to low-mass WIMPs ($\lesssim$10 GeV/$c^2$). \\ 

\section{Acknowledgements}
This work was supported in part by the National Science Foundation.
We thank Matt Pyle, Xavier Defay and Nicholas Zobrist for their help and useful discussion in performing this work.


\end{document}